\documentclass{elsart}
\journal{Philosophy Now}

\usepackage{graphicx,amssymb}

\begin{document}

\begin{frontmatter}

\title{A thought experiment on consciousness\thanksref{ded}}
\thanks[ded]{Dedicated to the memory of my grandfather Giulio-Fiore.}
\author{Germano D'Abramo}
\address{Istituto di Astrofisica Spaziale e Fisica Cosmica,\\ 
Area di Ricerca di Tor Vergata, Roma, Italy.\\
{\tt E-mail: Germano.Dabramo@rm.iasf.cnr.it}}

\date{\small First draft: August 2003}

\begin{abstract}

The Mind-Body Problem, which constitutes the starting point for a large
part of the speculations about consciousness and conscious experience, can
be re-stated in an equivalent way, using the `brain duplication' argument
described in this paper. If we assume that consciousness follows from a
peculiar organization of physical matter and energy, namely that it does
not transcend physical reality, then the brain duplication argument gives
a possible interesting physical characterization of the mind: namely, a
sort of extensive interdependence of the brain with the whole surrounding
physical world in giving rise to consciousness.

\end{abstract}


\end{frontmatter}


\section{Introduction}

One of the most fundamental problem in dealing with conscious experiences
and consciousness is the following: if I am able to completely describe
the physical state of my brain (conceding that all the physics necessary
to such description is already known), may I safely say to have completely
described my mental state, my subjective experience too? The point is that
my subjective experiences, like for example those of pain, joy or smell
(generally referred to as {\em qualia}), seem not to get exhausted in a
physical-functional description of my cerebral states, even in the most
complete description we are able to imagine to. Actually, the description
of the physical processes which take place in my brain, when I experience
pain for example, seems not to be a complete description of my subjective
experience of pain; at most, it seems to be only a complete description of
the cerebral states of my brain {\em during} my pain experience.

In other words, it seems that a barrier, impassable to every physical
theory, forbids any complete objective description of subjective
experience, or, at least, every complete objective description of a
subjective experience simply does not include the subjective experience
itself.  The objective description and the subjective experience seem to
belong to different and `orthogonal' dimensions, the {\em outside} and the
{\em inside}.

What I have described above briefly summarizes the well-known {\em
Mind-Body Problem}, the main ingredient of the philosophical
investigations of the mind and a thorn in the side of physicalism, namely
of those who believe in a complete reduction of consciousness to peculiar
physical processes of the brain (for accessible and exhaustive reviews
of the Mind-Body Problem see, for example, Nagel~\cite{na} and
Chalmers~\cite{dc1}).

In this paper I provide an equivalent formulation of the Mind--Body
Problem, which I will call the `brain duplication' argument, and I will
show that if we assume that consciousness is in any case a physical
process which takes place in the physical world, in the most general sense
of these terms, namely that it does not transcend physical reality, then
the human brain, in giving rise to the mind, might be characterized by the
astonishing property of an extensive interdependence with the {\em whole}
surrounding physical world.

\section{The `brain duplication' argument}

For the sake of thought experiment, let us suppose that we manage to
create an exact, physically identical duplicate of my brain, as it is in
this precise moment. Actually, it does not matter whether we do not known
operatively how to do it. The point is that since at least a brain exists
physically, then nothing forbids us to imagine an identical physical
duplicate of it, as well as nothing forbids us to imagine an identical
physical duplicate, atom by atom, of the sheet of paper on which these
words are written, just for the fact that such sheet of paper exists, even
if probably we will never be able to do the actual duplicate.

Now, just after the creation of this duplicate, how would my own
consciousness react? As everyone of you can experience directly, one of
the leading characteristics of consciousness is the perception of its own
uniqueness, the uniqueness of oneself and of its own conscious experience,
and, in addition, the perception of the persistence of such uniqueness (I
feel to be myself also in different periods and different places).  
Therefore, if I were able to exactly duplicate my brain, would my own
consciousness change? And if it changes, how does it change? Would I feel
to be here, where I was before the duplication, but at the same time would
I feel to be there, where my brain duplicate is now?

I believe that the most natural answer for everyone is that I will
continue to be myself as I was before the duplication. But then, {\em who}
or {\em what} is my exact physical duplicate? If the two brains are
physically identical and their consciousnesses are different, in what do
they differ? Here two possible approaches to the problem are presented.

\section{Non-physical explanation}

Consciousness is not physically reproducible in the reality. Namely it
does not depend on physical reality and it is in some sense `outside' it:
thus, distinct conscious experiences and consciousnesses may even be
attached to two physically identical brains, or only one of these brains
may be conscious (and the other one may not; the duplicate brain, for
example, might be a so-called {\em zombie}). If it is so, there is not
much to do; as a matter of fact, consciousness would constitute a {\em
prime} and {\em alien} property, to be added to the rest of the physical
properties of the brain.

However, such hypothesis seems to be scientifically frustrating and, after
all, not particularly reasonable. I guess that not many researchers would
honestly feel up to deny any link between consciousness and the physical
world (even if such link is not completely clear from a scientific point
of view). Even if we assume that the existence of the brain is a necessary
but not sufficient condition for the presence of consciousness, the latter
must necessarily have a physical interaction with the brain, otherwise it
would not even make sense to talk about brain as necessary condition,
without speaking about the substantial amount of data achieved nowadays in
the neurosciences on the {\em neural correlates} of consciousness.  
Moreover, matter can act on mind and consciousness, as any physical and
chemical interferences on our state of consciousness can easily
demonstrate. So, matter, mind and consciousness have to speak the same
`language'. And therefore consciousness must be a physically
characterizable, a physically tractable entity.

Such physical entity might be completely internal or external (partially
or totally) to the brain. In the first case we just would have that the
exact physical reproduction of one's brain would give {\em tout court} the
consciousness of that person (and hence, the brain duplication dilemma).
If, on the contrary, this entity were external, it might be one and
unique, or there might be many, one of them corresponding to each
consciousness currently existing in the world. Yet, this further
distinction is not important in our case: plausibly, two physically
identical brains would always interact in the same way with the same
external physical agent (like two equally tuned radio receivers `interact'
always with the same radio station, although there are a lot of different
radio frequencies in the air), and again the brain duplication dilemma
would be left untouched.  However, I believe that now this dilemma is
ready to be tackled with the possible physical explanation described in
the next section.

\begin{figure}[t]
\centerline{\includegraphics[width=11cm,angle=0]{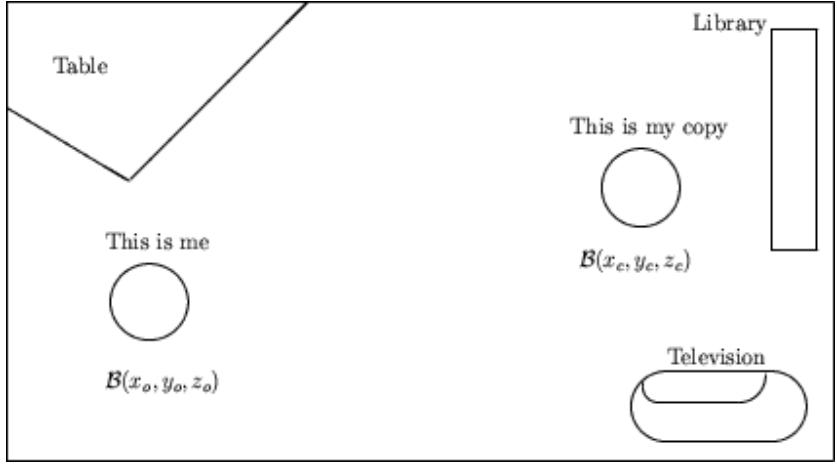}}
\caption{A na\"{\i}ve sketch of the argument described in the text.}
\label{fig1}
\end{figure}

\section{A possible physical explanation}

Let us suppose instead that consciousness depends on physical reality,
namely on the peculiar organization of physical matter and energy, in the
most general sense of these terms, and also let us do not exclude the
possibility, suggested in the previous section, that consciousness
originates not only in the physics of the brain but also through the
interaction with an external physical process.

A possible explanation of the seemingly paradoxical picture originated in
the brain duplication argument is that (the evolution of) the cerebral
processes involved in the rise of consciousness might physically depend
not only on the physics of the brain itself but actually also on {\em all}
the things which physically surround such brain, in the sense that it
depends on the organization of {\em all} the surrounding physical matter
and energy.

In such framework, the physically exact copy of my brain, which is in the
region of the space $\mathcal{B}(x_c,y_c,z_c)$, interacts with all the
surrounding physical world, and thus also with my original brain which is
in the region of space $\mathcal{B}(x_o,y_o,z_o)$. Similarly, my own original
brain, which is in the region of space $\mathcal{B}(x_o,y_o,z_o)$, interacts
with all the surrounding physical world, and thus also with the physically
exact copy of my brain, which is in $\mathcal{B}(x_c,y_c,z_c)$. Therefore, it
is clear (see fig.~\ref{fig1}) that the {\em boundary conditions} for the
two identical brains {\em are not identical}, and the evolution of
cerebral states, as well as consciousness, might not be identical
(although the memory of all the past experiences might be the same in
every details for both brains). A similar conclusion on the possibility of
an extensive physical interdependence of the human brain with the whole
surrounding physical world in giving rise to mind has been drawn in
another thought experiment (D'Abramo~\cite{dab}) but through a different
approach involving the notion of algorithmic complexity.

Note that any finite version of this approach, i.e.~that the suggested
physical dependence is only on a finite portion of the surrounding
physical world, does not change much the point. As a matter of fact, the
brain+`finite portion of physical world' must be a physically isolated
system, i.e.~a totally isolated system. For if it were not, two such
systems within two different environments would eventually lead to the
same dilemma on consciousness as before, since the same consciousness will
eventually experience different environments at the same time. Hence, even
in any finite approach, the finite portion of the surrounding physical
world is in fact {\em all} the surrounding physical world. Now, whether
there are spatial limits to be understood in the word ``{\em all}\,'' used
before, and whether they are posed by the concept of {\em visible
universe} and finite speed of light, the discussion of this is beyond the
scope of the present note.

Thus, the brain might be far from being a semi-closed box, opened only to
the five senses, and the common perception that our brain does not
continuously interact at a deeper physical level with the rest of the
physical world might be simply wrong. As a loose analogy, if the physical
reality were the water of a sea, our brain would not be like a submarine
guided by sonar, rather it would be more like a soaked sponge.

So far, we have never made mention to {\em how} consciousness rises, or to
{\em which} are the specific physical `mechanisms' at the basis of
consciousness, but this was not the topic of our paper, other than being a
very complicated and long-standing issue (irresolvable at the moment, I
think). Rather, I have proposed a possible physical characterization of
consciousness and mind: the uniqueness of conscious experience and
consciousness might depend also on the {\em whole} surrounding physical
world, in the sense of the organization of its physical matter and energy.
But, if this were the case, it would be very strange if the rise itself of
consciousness and mind did not depend on the extensive interaction of the
brain with {\em whole} surrounding physical world.

By the way, if we do not accept the picture of the extensive physical
interaction between brain and the whole surrounding physical world, but
want to maintain the dependence of mind and consciousness on physical
reality, and thus that mind and consciousness completely arise from the
physical organization of the brain alone, all this would inevitably result
in another type of `extensive interdependence', a sort of non-locality: as
a matter of fact, after the brain duplication, the very same mind (the
very same person)  could be in two very distant places at the same time.

I believe that talking about the necessary physical mechanisms responsible
for the suggested extensive interaction, and actually talking about the
detailed physical dynamics and evolution of such interaction, is premature
at the moment. It might be said (and it was actually said already) that
the human brain may be assimilated to a complex dynamical system,
extremely sensitive to all the surrounding physical conditions (in such
case we should mention the so-called {\em deterministic chaos}; see for
example Newman~\cite{new}), or that quantum mechanics may be deeply
involved (many people have proposed various quantum mechanisms to explain
consciousness, so to compile a complete list of references on this topic
is hopeless; see for example Bohm~\cite{boh} and references therein), or,
most probably, that a new kind of physics is needed. However, although
such aspect of the problem is obviously essential in the study of the
origin of consciousness, it is secondary in the present context.

\section*{Acknowledgments}

I wish to thank Alessandro Silvestrini for having brought to my attention
the `brain duplication' argument, during a conversation around a coffee
table. By the way, the name of the Caf\`e was `Mir\'o', thus for the sake
of joke, I suggest to call the argument of this paper `the Mir\'o
hypothesis'. I am also grateful to William A.~Adams for insightful
comments on this paper. I wish to thank Herbert F.~Muller for having
posted a previous version of this paper on the Karl Jaspers Forum Website.

\end{document}